\documentclass[aps,a4paper,showpacs,twocolumn]{revtex4}

\usepackage{epsfig}
\usepackage{amsmath,amssymb,color}
\usepackage[english]{babel}

\parskip=\medskipamount

\newcommand{\eq}[1]{(\ref{#1})}
\newcommand{\fig}[1]{Fig.\ref{#1}}

\newcommand{\be}{\begin{equation}}
\newcommand{\ee}{\end{equation}}

\newcommand{\barr}{\begin{array}}
\newcommand{\earr}{\end{array}}

\newcommand{\beqn}{\begin{eqnarray}}
\newcommand{\eeqn}{\end{eqnarray}}

\newcommand{\bs}{\begin{subequations}}
\newcommand{\es}{\end{subequations}}

\newcommand{\bw}{\begin{widetext}}
\newcommand{\ew}{\end{widetext}}

\begin{document}

\title{Islands of stability in motif distributions of random networks}

\author{M.V. Tamm$^{1,4}$, A.B. Shkarin$^2$, V.A. Avetisov$^{3,4}$, O.V. Valba$^{4,5,6}$,
S.K. Nechaev$^{4,5,7}$}

\affiliation{$^1$Physics Department, Moscow State University, 119992, Moscow, Russia \\
$^2$Department of Physics, Yale University, 217 Prospect Street, New Haven, CT 06511, USA \\
$^3$N.N. Semenov Institute of Chemical Physics of the Russian Academy of Sciences, 119991, Moscow,
Russia \\ $^4$Department of Applied Mathematics, National Research University Higher School of
Economics, 101000, Moscow, Russia \\ $^5$Universit\'e Paris--Sud/CNRS, LPTMS, UMR8626, B\^at. 100,
91405 Orsay, France \\ $^6$Moscow Institute of Physics and Technology, 141700, Dolgoprudny, Russia
\\ $^7$P.N. Lebedev Physical Institute of the Russian Academy of Sciences, 119991, Moscow, Russia}


\begin{abstract}

We consider random non-directed networks subject to dynamics conserving vertex degrees and study
analytically and numerically equilibrium three-vertex motif distributions in the presence of an
external field, $h$, coupled to one of the motifs. For small $h$ the numerics is well described by
the "chemical kinetics" for the concentrations of motifs based on the law of mass action. For
larger $h$ a transition into some trapped motif state occurs in Erd\H{o}s-R\'enyi networks. We
explain the existence of the transition by employing the notion of the entropy of the motif
distribution and describe it in terms of a phenomenological Landau-type theory with a non-zero
cubic term. A localization transition should always occur if the entropy function is non-convex. We
conjecture that this phenomenon is the origin of the motifs' pattern formation in real evolutionary
networks.

\end{abstract}


\maketitle

Studying of complex networks constitutes a rapidly developing interdisciplinary area
\cite{barab,dorogov}, which unites investigation of various types of experimentally observed
biological \cite{barab2}, social \cite{skott,newman,jackson0} and engineering \cite{bollobas}
networks, as well as artificial random graphs constructed by various probabilistic techniques
\cite{erdos,barab3, krapiv,dorogov2,vicsek}. Many statistical properties of network, including the
vertex degree distribution, clustering coefficients, ``small world'' structure \cite{watts} and
spectra of adjacency matrices \cite{goh} have been extensively treated.

One particular topological characteristic, which seems to be very instructive in providing detailed
information about local network topology, is the distribution of small subgraphs (motifs). The
presence of motifs of specific type is tightly linked to the network function. For example, the
transcriptional regulatory networks have excess of auto-regulation loops with respect to randomly
connected graphs with the same vertex degree distribution \cite{Balazsi2005, Shen-Orr2002}. Some
authors \cite{Conant2003} connect the prevalence of specific motif types in the transcription
factor networks of \emph{E.coili} and \emph{S.cerevisiae} with a network evolution, which might
lead to self-consistent optimal circuit design \cite{Maayan2005, Csete2002, Mangan2003}. It is
known, that protein interaction networks have many short cycles and completely connected subgraphs
\cite{Wuchty2003}. This can be thought as a feature necessary for a signal transduction by
feed-back loops \cite{Maayan2005}. These and many other examples clearly demonstrate that the
existence of specific motifs in networks strongly correlate with the network function.

The statistical analysis of motifs distribution (MD) in various naturally observed directed
networks demonstrates \cite{alon,alon1} that the simplest three-vertex motifs (triads) of oriented
graphs can be split into four broad \emph{superfamilies}, and the networks within the same
superfamily tend to have similar functions. However, to the best of our knowledge, still there is
no common opinion about the mechanism behind the separation of MDs into a particular preferred
stable state. In this letter we put forward a hypothesis which may give at least a partial answer
to this question.

The basic idea of our approach is to notice that a MD gives an essentially coarse-grained
description of a network: one integrates out many internal degrees of freedom. Therefore, there
exist an entropy corresponding to each particular motif distribution. If some distribution is
entropically advantageous, it will occur more often and acts as an effective trap for the network
dynamics. Such favorable distributions are the \emph{islands of stability} in a sea of all MDs, as
conjectured in \cite{shkarin}.

We consider here statistical properties of simplest motifs -- the three-vertex subgraphs or triads
-- in $N$-vertex random networks. The microscopic configuration of the network is completely
defined by $N(N-1)/2$ Boolean variables denoting presence/absence of edges, while the MD is
described by a vector ${\bf M}$ whose elements are the numbers of triads $\{M_0,M_1,...\}$, or,
more conveniently, by vector ${\bf c}$ of triad concentrations, $\{c_0,c_1,...\}$, $c_i = M_i/M$,
where $M=N(N-1)(N-2)/6$ is the total number of triads in the network \footnote{In \cite{alon} the
authors use renormalized densities of the motifs. In our notation, the components of their motif
vector ${\bf Z}$ are equal to $Z_i=(c_i-c_i^0)/\chi_i=\Delta c_i/\chi_i$, where $c_i^0=c_i(h=0)$,
and $\chi_i$ is the susceptibility of $i$th concentration to an external field $h_i$ coupled to it,
taken at $h_i=0$. This linear shift does not change the results presented in our work.}. For
non-directed graphs there are 4 different triad types, shown in the \fig{fig:1}, while for directed
graphs there are 16, however not all of them are independent (see below).

Studying the entropy of a network as a function of MD from the first principles seems to be an
overwhelmingly difficult task. Instead, we model a network in an auxiliary external field, ${\bf
H}$ coupled to the triad distribution, that is, we consider an ensemble of networks with the
partition function $W({\bf H})=\sum_X^{\prime} e^{-{\cal H}(X)/T}$ with ${\cal H}(X) = -{\bf
H}\,{\bf M}(X)$, where the sum runs over all microscopic configurations $X$ and $^{\prime}$
designates conditions imposed on the configurational space (e.g., the conservation of the network
degree distribution). Clearly, the unbiased case of equiprobable network configurations corresponds
to ${\bf H}={\bf 0}$. This model is essentially athermic: in the absence of the external field the
partition function of the network is purely a combinatorial object with no internal interactions
and no temperature dependence (compare this, e.g., with athermic liquid mixtures or
non-self-intersecting random walks \cite{cloiz}), it seems therefore natural to introduce
normalized dimensionless field ${\bf h}={\bf H}/T$, which is the only external variable governing
the behavior of the model. This approach, reminiscent of the biased molecular dynamics used, for
example, in \cite{mueller,deng} allows us, by varying ${\bf h}$, to skew the motif distribution and
thus sample the states of a network which are otherwise unaccessible. As a result, we obtain a full
free energy landscape of a network as a function of motif distribution. To equilibrate a network in
an external field we permute links as shown in \fig{fig:1}a, preserving the node degree
distribution \cite{sneppen} and use the Metropolis algorithm to accept or reject single steps.

\begin{figure}[ht]
\epsfig{file=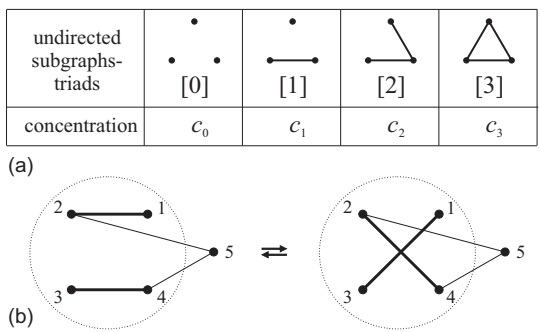,width=7cm} \caption{a) Possible triads in a non-directed network; b)
Single link permutation: links $(12)$ and $(34)$ are removed, and links $(13)$ and $(24)$ are
created. Triad $\{135\}$ goes from type [0] to type [1], triads $\{125, 345\}$ -- from type [2] to
type [1], and triad $\{245\}$ -- from type [2] to type [3]: three new triads of type [1] and one
triad of type [3] are created instead of three triads of type [2] and one of type [0], compare to
Eq.\eq{eq:1}.}
\label{fig:1}
\end{figure}

For ${\bf h}={\bf 0}$ the system lives in the largest entropic basin corresponding to some
equilibrium distribution of motifs. As $|{\bf h}|$ is increased, the motif distribution gets
gradually more skewed. In the limit $|{\bf h}| \to \infty$ the entropic effects become irrelevant,
and the network approaches the state with the largest possible value of the ``energy'', i.e. of the
product ${\bf h}\,{\bf M}$. Depending on the particular shape of the entropy function, the motif
vector ${\bf M}$ can be either a smooth function of external field ${\bf h}$, or it can undergo
abrupt jumps at some particular values of ${\bf h}$ resembling first-order phase transitions.

Here we look at the simplest case of the triad distribution in undirected networks. The
randomization procedure \cite{sneppen} consists of repeated permutations of randomly chosen pairs
of links -- see \fig{fig:1}. Each permutation changes the number of triads of different types: four
vertices encircled in \fig{fig:1} belong to $4 + 6(N-4) + 2 (N-4)(N-5)$ different triads, of which
(i) 4 consist of three vertices constructed from the set $(1234)$, all of them are of type [1] and
do not change in the elementary step; (ii) $6(N-4)$ triads including two vertices from $(1234)$ and
one external vertex, some of them change as a result of a permutation (we call these changes
``elementary reactions'' in what follows, not to be confused with ``permutation steps'' which,
generally speaking, consist of many simultaneous elementary reactions), (iii) $2(N-4)(N-5)$ triads
including only one of the vertices from the set $(1234)$, they do not change in a permutation. A
direct check shows that for an undirected network there is only one possible type of elementary
reaction which changes the numbers of the triad types:
\be
[0]+3[2] \rightleftarrows [3]+3[1],
\label{eq:1}
\ee
When a new [0]-triplet is formed, it always coincides with formation of 3 triplets of type [2], and
elimination of one [3]-triplet and three [1]-triplets (see \fig{fig:1}). Equation \eq{eq:1} sets a
connection between the time derivatives of the triads concentrations:
\be
3\dot{M}_0(t)=\dot{M}_2(t)=-3\dot{M}_3(t)=-\dot{M}_1(t),
\label{eq:conserv}
\ee
where $\dot{M}_i\equiv \frac{dM_i}{dt}$ ($i=0...2$). Since only one triad concentration is
independent, the undirected network is effectively 1D in terms of triads distributions. This means
that there exist three independent combinations of $c_0,...,c_3$, which are conserved under
\eq{eq:conserv}, one can choose them as
\be
\begin{array}{l}
I_1=\sum_{i=0}^{3} M_i = M; \quad I_2=\sum_{i=0}^{3} i M_i = 3pM; \medskip \\ I_3=\frac{1}{2}(M_0 +
M_3),
\end{array}
\label{eq:laws}
\ee
$I_1$ and $I_2$ are, respectively, the total number of triads, $M$, and links, $3pM$, and $p$ is
the fraction of bonds (links) formed in the system. The invariance of $I_3$ follows from the
conservation of the vertex degree distribution (see the supplementary materials to \cite{alon}). We
parametrize the one-dimensional evolution of motifs by a variable $m=\frac{1}{2}(M_3-M_0)$, or
$c=m/M$. The external field $h$ coupled to $m$ is also one-dimensional, and the dimensionless
network energy in the external field reads
\be
E = - h\, m = - \frac{1}{2} h (M_3-M_0).
\label{eq:energy}
\ee
According to the detailed balance rule, in an equilibrium network subject to energy $E$, for the
probabilities of forward, $p_+$, and backward, $p_-$, permutation steps, one has $p_+/p_- =
e^{-\Delta E}$, where $\Delta E$ is the overall energy change due to all elementary reactions
invoked by this permutation step. In the Metropolis algorithm this is achieved by accepting the
reaction with probability $1$ if it decreases $E$, and with probability $\exp\{-\Delta E\}$
otherwise.

Since each permutation implies many simultaneous elementary reactions \eq{eq:1}, the reactions are,
generally speaking, not independent. A natural first approximation is, nevertheless, to neglect
correlations between them and apply the law of mass actions \cite{lawmassact,landau} to the
chemical kinetics described by Eq.\eq{eq:1}. It is straightforward to obtain the implicit
$c(h)$-dependence
\be
K \equiv \frac{p_+}{p_-}= e^h = \frac{c_3 c_1^3}{c_0 c_2^3} = \frac{(A + c)(2- 3p - A + 3c)^3}{(A -
c)(3p - 1- A -3c)^3},
\label{mass}
\ee
where $A=I_3/M$ (see \eq{eq:laws}), and $K=e^{h}$ is the equilibrium reaction constant ($K=1$ for
$h=0$).

\begin{figure}[ht]
\epsfig{file=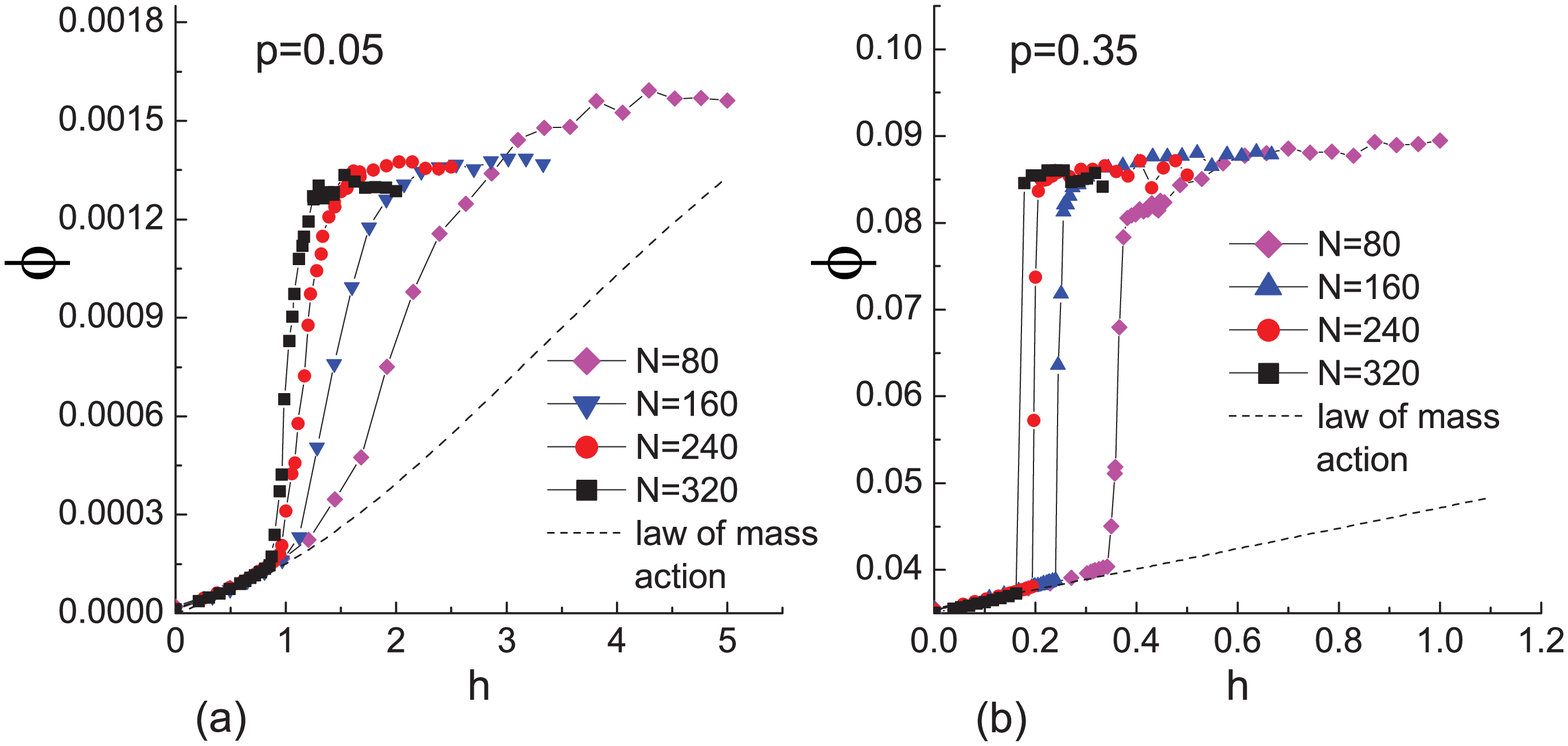,width=8cm} \caption{(Color online). The motif distribution $\phi(h)=
c - c(h=0)$ in ER networks with $p=0.05$ (a) and $p=0.35$ (b), re-equilibrated at different $h$.
Solid lines -- numerical results for networks of sizes $N=80,$ (magenta diamonds) $160$ (blue
triangles), $240$ (red circles), and $320$ (black squares), dashed line -- predictions of the law
of mass action \eq{mass}.}
\label{fig:2}
\end{figure}

The mean-field equation \eq{mass} is applicable to networks with any degree distribution. In what
follows we concentrate, though, on a particular example of a conventional Erd\H{o}s-R\`{e}nyi (ER)
network \cite{erdos} with bond formation probability $p$. The triad concentrations in this case are
given by:
\be
\begin{array}{l}
\bar{c}_0=(1-p)^3;\, \bar{c}_1=3(1-p)^2p;\, \bar{c}_2=3(1-p)p^2;\, \bar{c}_3=p^3; \medskip \\
\bar{A}= \frac{1}{2}(\bar{c}_0+\bar{c}_3) = \bar{I}_3/M = (p^3 + (1-p)^3)/2
\end{array}
\label{eq:5}
\ee
All concentrations, as expected, satisfy \eq{mass} with $K=1$. Starting with different $p$, we
perform a randomization at non-zero fields $h=\ln K > 0$. The typical resulting dependences
$\phi(h)=c(h)-c(h=0)$ are shown in \fig{fig:2}a,b for networks created at $p=0.05$ (a) and $p=0.35$
(b) for various network sizes, $N$. The dashed lines show the $\phi(h)$ dependence as given by
solving \eq{mass} with $A=\bar{A}$ given be \eq{eq:5}. At high $h$ the $\phi(h)$ curve saturates
due to depletion of triads of type 2 with growing $c$. In the vicinity of $h=0$ the numerical
results are in good agreement with the law of mass action (LMA) \eq{mass}, while for larger $h$, a
sharp change in $\phi(h)$, not predicted by the LMA, is registered. This discrepancy between theory
and numerical experiment is, of course, due to the correlations between elementary reactions
constituting a single permutation step. To check this, we study\cite{supp} the distribution of the
number of forward and backward elementary reactions constituting a single permutation step and show
that while for small $h$ this distribution is nearly Gaussian (as one expects for independent
elementary reactions), in large fields the distribution acquires a peculiar bimodal shape signaling
strong correlations between elementary reactions.

For large $p$ the transition between two regimes is very narrow, being, as we shall see, the
first-order phase transition. In \cite{supp} we show that in this regime a hysteresis typical for
first-order transitions is observed. For smaller $p$ the transition is less narrow due to larger
fluctuations, although its width decays with increasing $N$. For $p=0.05$ the difference between
the specific free energies of two competing phases is relatively small, and both phases could
coexist within the whole transition region (compare to the coexistence of two phases in the
liquid-liquid 1st order phase transition reported in \cite{stanley}).

To verify this, we have measured (see \cite{supp}) the distribution of links in the network with respect to
the number of triangles (type [3] subgraphs) they belong to. The clear bimodality of the distribution allows
us to conclude that  that there is indeed a phase coexistence in the system: the links in almost
triangle-free (low-$h$) phase participate only in few triangles, while the links in the triangle-rich
(high-$h$) phase typically participate in many triangles.

We describe the observed transition via a phenomenological Landau-type theory \cite{landau} with
$\phi$ as an order parameter. Since there is no $\phi \leftrightarrow - \phi$ symmetry in the
problem, the Landau expansion of the free energy is expected to include both odd and even powers of
$\phi$, so up to the 4th order term we have
\be
\begin{array}{l}
H = H_0- M h\, \phi; \medskip \\
H_0/M = \frac{\chi}{2} \phi^2 - \frac{b(N,p)}{3} \phi^3 + \frac{g(N,p)}{4} \phi^4 + o(\phi^4)
\end{array}
\label{eq:9}
\ee
Here $H_0$ is a purely combinatorial (i.e., temperature and field--independent) free energy of a
network with given motif distribution in the absence of external field, by definition it has a
minimum at $\Delta c=0$. The zero-field susceptibility $\chi$, according both to LMA (see
\eq{mass}) and computer simulations at small $h$, is $N$-independent and can be computed in a
straightforward way $\chi\equiv \frac{\partial^2 H(\Delta c)}{\partial (\Delta
c)^2}\Big|_{h=0}=\frac{\partial h(c)}{\partial c}\Big|_{h=0}$ giving
\be
\chi = \frac{1}{\bar{c}_0} + \frac{9}{\bar{c}_1}+ \frac{9}{\bar{c}_2}
+\frac{1}{\bar{c}_3}=\frac{1}{p^3(1-p)^3},
\label{chi}
\ee
where the last equality is true only for ER networks. In turn, higher order coefficients $b(N,p)$
and $g(N,p)$ are expected to be $N$-dependent.

The structure of the Hamiltonian \eq{eq:9} allows for a first order phase transition. Indeed, the
equilibrium value, $\phi$, is defined by minimizing $H(\phi)$ in \eq{eq:9}, and is given implicitly
by equation
\be
\chi\,\phi - b\,\phi^2 + g\,\phi^3=h
\label{eq:10}
\ee
For $b^2<3g \chi$ this equation has a single solution for any $h$ (i.e., the free energy $H_0$ is
always convex), but for $b^2>3g \chi$ there exist a region with three solutions, which correspond
to two competing minima of the free energy and one unstable maximum in between. In the
thermodynamic limit $N \to infty$ the transition occurs when the values of $H$ match in the minima,
while in smaller systems the fluctuations smear the transition.

\begin{figure}[ht]
\epsfig{file=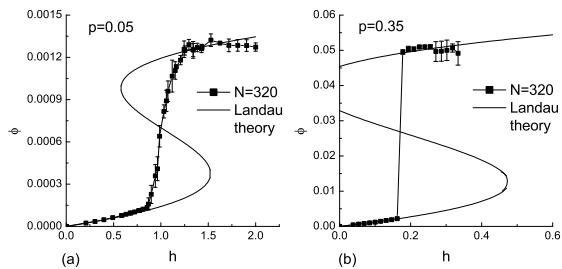,width=8cm} \caption{Comparison of the numeric dependence
$\phi(h)$ to the solution of the mean-field equation \eq{eq:10} a) numerical results for $N=320,
p=0.05$ (squares) and the best fitting Landau theory with $\chi=p^{-3}(1-p)^{-3}=9330$,
$b=1.72\times 10^7$, $c=8.45 \times 10^9$; b) numerical results for $N=320, p=0.35$ (squares) and
the best fitting Landau theory with $\chi=p^{-3}(1-p)^{-3}=85$, $b=4.46\times 10^3$, $c=5.7 \times
10^4$.}
\label{fig:3}
\end{figure}

The numerics for the $\phi(h)$-dependence is well fitted both  for $p=0.05$ and $p=0.35$ with the
ansatz \eq{eq:10} -- see \fig{fig:3}. Note that while at $p=0.35$ the transition looks as a
discontinuous jump of the order parameter at a point defined by a Maxwell rule,  at $p=0.05$ the
behavior in the transitional region is given by a linear combination of two competing phases
(compare to \cite{stanley}).

Importantly, the transition points shifts to lower $h$ with increasing size of the network. The
detailed analysis of the limiting behavior for $N \to \infty$ and its impact on the coefficients of
the Landau free energy \eq{eq:9} goes beyond the scope of this paper, but se show that the
transition \cite{supp} value of $h$ seems to decay as $N^{-\alpha}$ with the system size $N$, where
$\alpha = 0.5 \pm 0.1$. In turn, the width of the transition decays more rapidly with $N$,
approximately as $N^{-1}$.

To summarize, we argue that a projection of microscopic (in terms of network nodes) onto
macroscopic (in terms of triad concentration) description gives rise to a nontrivial dependence of
the entropy on a given motif distribution. In the simplest case of non-directed network considered
here, the macroscopic description is effectively one-dimensional. In the presence of an external
field, $h$, the equilibrium value of the motif concentration, $\Delta c=c(h)-c(h=0)$, is determined
by the balance of energy associated with $h$, and the entropy originating from a mapping of
microscopic to macroscopic description. If the entropic landscape is concave, a phase transition
into a state with highly-skewed motif distribution occurs at some critical $h_{\rm c}$. This
transition, observed numerically for ER networks, violates the law of mass action due to
correlations between elementary reactions \eq{eq:1} in strong fields.

How general is this transition phenomenon? To check this, we modified the elementary permutation
rules allowing an edge connecting two arbitrary vertices $(i,j)$, to be switched to some other pair
$(i,k)$ ($k\neq i,j$). Under this dynamics the nodes degrees are not conserved, and the integral
$I_3$ in \eq{eq:laws} is absent. Accordingly, the dynamics in the motif space becomes effectively
two-dimensional with elementary reactions
$$
2[2] \rightleftarrows [1] + [3]; \quad 2[1] \rightleftarrows [0] + [2]
$$
However, application of an external field ${\bf h}$ (which in this case is a 2D-vector) still leads
to a transition to a localized motif distribution\cite{avetisov_and_friends}. Another example of a
similar transition, which corresponds in the notations of this paper to the dynamics with random
creation and elimination of bonds with the external field ${\bf h} ={0,0,h}$ coupled to triangles
(the motif parameter space is in this case 3-dimensional) was first observed in \cite{Strauss} and
discussed theoretical in \cite{Newman}, note also the conjugate effect of bimodality in the
triangle parameter estimates in exponentially distributed networks reported in \cite{jackson}. In
\cite{supp} we provide preliminary data on networks with fixed scale-free distributions of edge
degrees, showing that similar localization transition happens there as well.

The phenomenon of localization of the motif distribution under external field into distinct
entropic traps is apparently wide-spread. We conjecture that stable motif profiles constituting
superfamilies \cite{alon} may correspond to such stability islands inherent to the complicated
underlying entropic landscape of a multidimensional motif space of directed networks.

The concept of entropically induced localization may be instrumental in various other fields.
Compare it, for example, with the celebrated Eigen model of biological evolution in the space of
heteropolymer sequences \cite{eigen}. There, the localization-delocalization phase transition,
known as the "error catastrophe", separates two states, where the genotype is localized in the
vicinity of a preferred pattern, and where it is completely random \cite{gall,peliti,slan}. The
transition occurs due to an interplay between the attraction to a point--like potential well and
the entropic repulsion from this well due to the exponential growth of the number of states with
increasing the Hamming distance from the well. In our case, a complimentary phenomenon occurs: the
entropic landscape of the system acts as a source of effective attractive traps, while the uniform
external field regulates the transitions between trapped states. It seems that trapping of a
complex system in stability islands due to a competition between selection and randomness, provides
a generic mechanism of localization in complex biological and social systems.

The authors are grateful to M. Kardar, P. Krapivsky, A. Mikhailov and K. Sneppen for valuable
discussions. This work was partially supported by the grunts ANR-2011-BS04-013-01 WALKMAT,
FP7-PEOPLE-2010-IRSES 269139 DCP-PhysBio, as well as by a MIT-France Seed fund and the Higher
School of Economics program for Basic Research.

\begin{appendix}

\section*{Supplementary materials}

\section{Distribution of backwards and forwards reactions.}

\begin{figure}[ht]
\epsfig{file=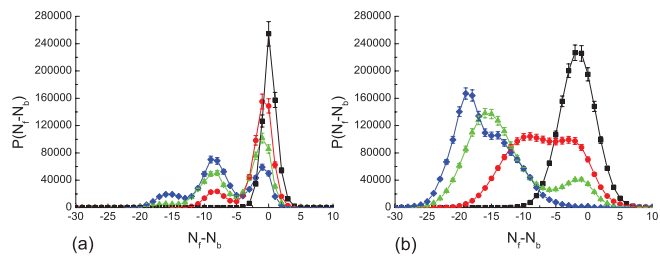,width=8.5cm} \caption{Distribution $P ({\cal N})$ for the ER networks
with $N=100$ and different values of $p$ and $\phi$: (a) $p=0.05$, $\phi = 10^{-4}$ (black
squares), $4.5 \times 10^{-4}$ (red circles), $8.5 \times 10^{-4}$ (green triangles), and $1.2
\times 10^{-3}$ (blue diamonds); (b) $p=0.35$ $\phi = 0.001$ (black squares), $0.011$ (red
circles), $0.028$ (green triangles), and $0.034$ (blue diamonds). The Y-axis scale is arbitrary.}
\label{fig:4}
\end{figure}

To study the correlations between elementary reactions we calculated the distribution $P({\cal N})$
of the difference, ${\cal N}={\cal N}_f-{\cal N}_b$, between ``forward'', ${\cal N}_f$ and
``backward'', ${\cal N}_b$, elementary reactions for \textit{attempted} permutation steps (i.e.,
regardless of whether it is accepted or rejected) in the skewed ER networks with different fixed
values of $\phi$. Note that since in the transition region it is difficult to tune exact value of
$\phi$ by choosing external field $h$, we have used a different way of pinpointing a network with
given $\phi$. Namely, a network with a desired $\phi=c^*-c(0)$ was obtained by equilibrating ER
network in a potential $H_c=\exp(\gamma|c-c^*|^2)$ with $\gamma$ sufficiently large to make the
fluctuations of $\phi$ (or $c^*$) negligible, so that the network becomes strongly localized in the
state with desired $\phi$. The results of this modelling are presented in the \fig{fig:4}.

Close to equilibrium, i.e. at $\phi \approx 0$ the distribution $P({\cal N})$ is shifted to the
left, but is still nearly Gaussian, meaning that different backward and forward reactions occur
independently and the LMA is still valid. However, as $\phi$ gets progressively larger, the
distribution $P({\cal N})$ becomes substantially non-Gaussian, developing a bimodal shape in the
transition region $0.01<\Delta c<0.05$. This indicates that the elementary reactions are no longer
independent, and all the permutations can be roughly divided into two classes: (i) those which do
not change much the motif distribution (the right peak in the distribution), and (ii) those which
lead to an essential reduction in the number of triads of type [3], pushing the system towards the
equilibrium motif distribution. As $\phi$ approaches the saturation value, the forward reactions
$[0]+3[2] \to [3]+3[1]$ get suppressed, as there are almost no subgraphs of type [2] left in the
system, and the backward reactions $[0]+3[2] \leftarrow [3]+3[1]$ become dominant.

\section{Hysteresis in the transition from triangle-poor to triangle-rich phase.}

\begin{figure}[ht]
\epsfig{file=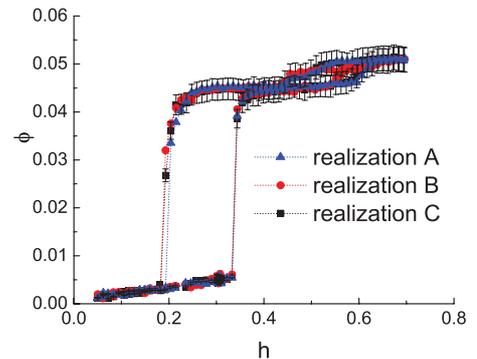,width=6cm} \caption{Hysteresis for $\phi(h)$. The curves A, B, C
correspond to different samples of adiabatic changes of the field $h$.}
\label{fig:hysteresis}
\end{figure}

If the transition from triangle-poor to triangle-rich state is a first order phase transition, one
expects a hysteretic behavior of $\phi(h)$. Such a behavior is indeed seen in the simulations for
three ER network samples (A, B, C) of size $N=100$ with $p=0.35$ shown in \fig{fig:hysteresis}. The
hysteresis has been recorded for the $\phi(h)$-dependence when the dimensionless field $h$ is
adiabatically increased from zero up to the maximal value $h=2$ and then adiabatically decreased
back to $h=0$.

\section{Distribution of links with respect to the number of triangles they are involved in}

\begin{figure}[ht]
\epsfig{file=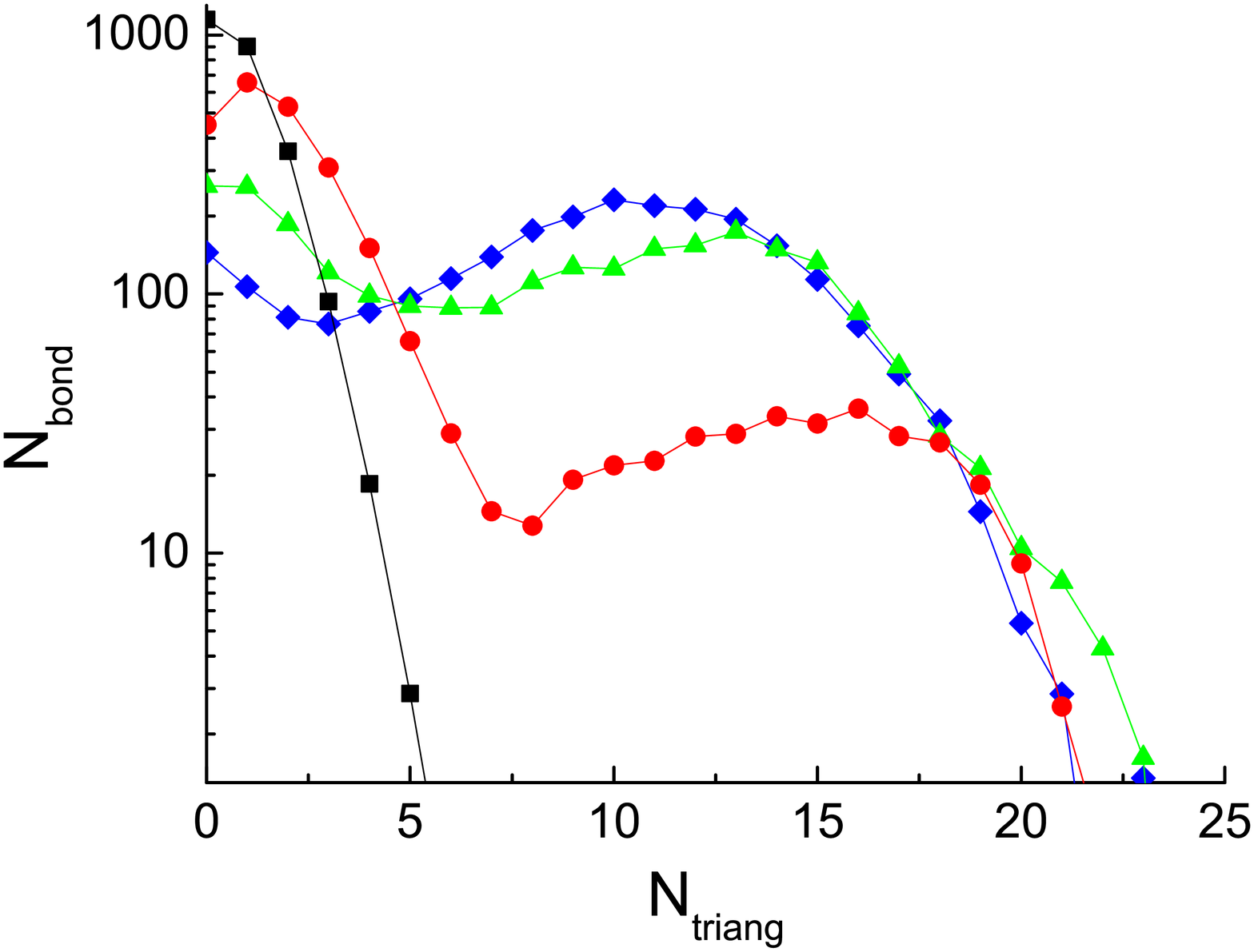,width=6cm} \caption{(Color online). Distribution of the links
in a network with $N=320, p=0.05$ with respect to the number of triads of type [3] (triangles) they
belong to for different values of the auxiliary field $h$: $h=0$ (black squares), $h=0.94$ (red
circles), $h=1.19$ (green triangles, $h=1.67$ (blue diamonds). The bimodality of the curve signals
the coexistence of the triangle-rich and triangle-poor phases.}
\label{fig:triangledist}
\end{figure}

In order to check that even for the ER networks with small fraction of bonds (e.g. $p=0.05$) the
transition from triangle-poor to triangle-rich phase is a first-order transition, we have measured
the distribution of links in the network with respect to the number of triangles (type [3]
subgraphs) they belong to. We expect this measure to correlate strongly with the phase particular
link belongs to, therefore, if there is a coexistence of two distinct phases, a bimodality in this
destribution is expected. This bimodality is readily observed in \fig{fig:triangledist} allowing us
to conclude that that there is indeed a phase coexistence in the system: the links in almost
triangle-free (low-$h$) phase participate only in few triangles, while the links in the
triangle-rich (high-$h$) phase typically participate in many triangles.

Integrating the first and second ``humps'' of the curve in \fig{fig:triangledist} one obtains that
roughly 2500 links are distributed between phases in proportion 2500:0 at $h=0$, 2200:300 at
$h=0.94$ (close to the lower end of the transition region), 1100:1400 at $h=1.19$ (close to the
upper end of the transition region), and 400:2100 at $h=1.67$ (above transition region). The latter
result may signal either that there is still some left-over triangle-poor phase above the
transition region, or, more probably, that the links distribution with respect to the number of
triangles in triangle-rich phase is rather wide, leading to some of its links been erroneously
identified as belonging to the triangle-poor phase. Thus, for small ($p=0.05$) and large ($p=0.35$)
density of links, the abrupt change in $\Delta c$ with changing $h$ in ER networks is a first order
phase transition from the low-$h$ phase with the MD close to the equilibrium ER network, to the
high-$h$ phase with a strongly skewed MD.

\section{Scaling as $N \to \infty$.}

\begin{figure}[ht]
\epsfig{file=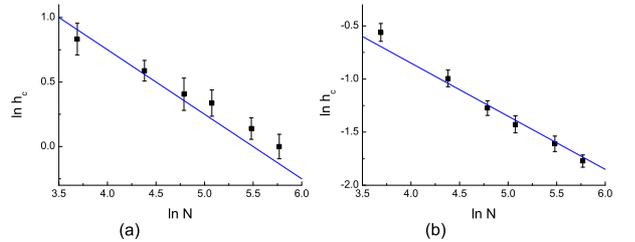, width=8cm} \caption{The $N$ dependence of the transition
point as measured by the largest slope of the $\phi(h)$ curve for Erd\H{o}s-R\'enyi networks with
(a) $p=0.05$ and (b) $p=0.35$. The blue straight lines on both figures have a slope of $-1/2$.}
\label{suppl:figscaling}
\end{figure}

\begin{figure}
\epsfig{file=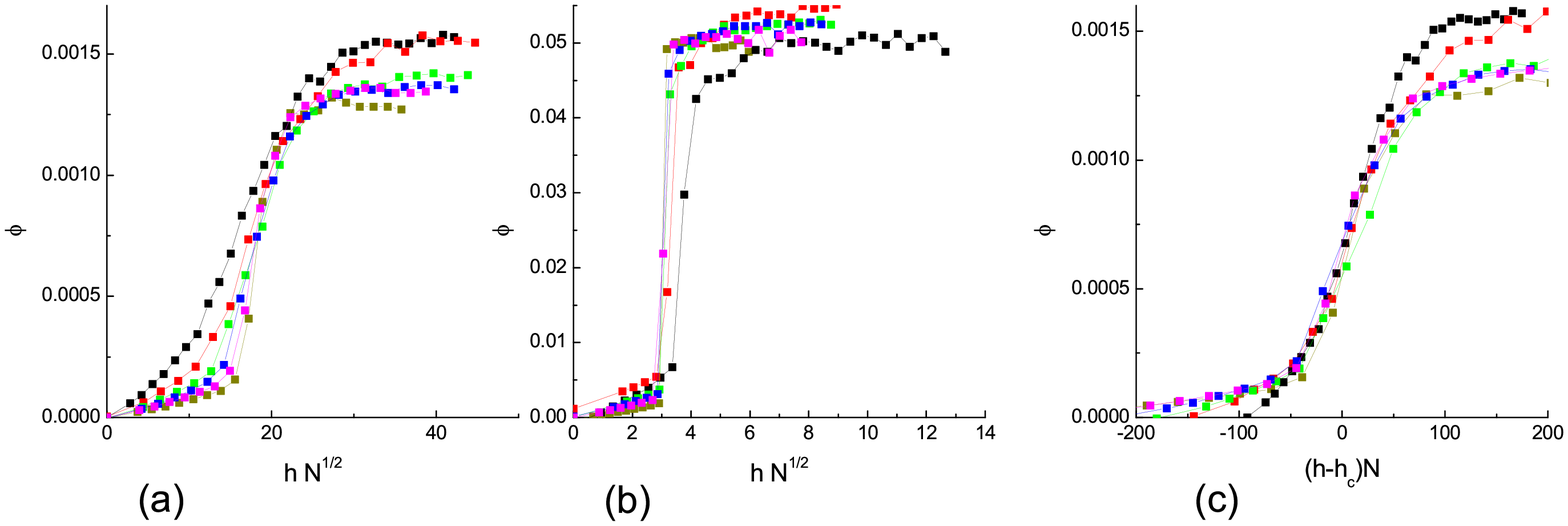, width=8.5cm} \caption{The $\phi(h)$ dependences in rescaled
coordinates for the networks with different $N$. (a), (b) collapse of the transition point in
$\phi(h \sqrt{N})$ coordinates for ER networks with $p=0.05$ (a) and $p=0.35$ (b). (c) the vicinity
of the transition point for $p=0.05$ plotted in $\phi(h N)$ coordinates. $N=40$ (black), $80$
(red), $120$ (green), $160$ (blue), $240$ (magenta) and $320$ (dark yellow) for all figures. .}
\label{suppl:collapse}
\end{figure}

As it is clearly seen from Figure 2 of the main text, the value $h_c$ at which the transition into
localized phase occurs decays with growing $N$. In fact, that is to be expected: the total number
of motifs in the network $M$ is proportional to $N^3$, while the combinatorial free energy (see eq.
7) should probably grow slower (e.g., proportional to the number of links $\sim N^2$).
\fig{suppl:figscaling} shows the dependence of the transition point (measured as the point with the
largest slope of the $h(c)$ dependence) on $N$ for different values of $p$. The results for both
$p$ are compatible with the hypothesis that $h_c \sim 1/\sqrt{N}$, although some divergences from
this simple dependence are observed: fitting the log-log plots with straight lines gives the slopes
of $\approx 0.43$ and $\approx 0.57$ for $p=0.05$ and $p=0.35$, respectively. As seen from
\fig{suppl:collapse}(a),(b) the transition points collapse reasonably well when redrawn in $\phi(h
\sqrt{N})$ coordinates. Meanwhile, the width of the transition for $p=0.05$ decays roughly as
$N^{-1}$ as shown in \fig{suppl:collapse}(c), meaning that even in rescaled coordinates with
constant transition point the width of the transition goes to zero with increasing $N$, once again
proving that the phenomenon we study is indeed a phase transition (in $p=0.35$ case the transition
width is negligible for all network sizes $N$ under consideration).

\section{Scale-free networks in external field.}

\begin{figure}[ht]
\epsfig{file=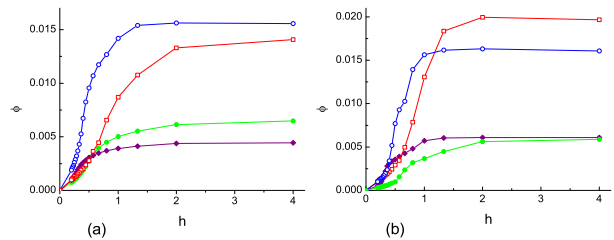, width=8cm} \caption{$\phi(h)$ dependence for networks with scale-free
node degree distribution $P(k) \sim k^{-\alpha}$, (a) $\alpha=2$, (b) $\alpha = 3$. Data for
networks with fraction of bonds equal to $p=0.1$ (filled symbols) and $p=0.2$ (open symbols) is
shown, network size $N$ is $50$ (squares), $100$ (circles), and $200$ (diamonds).}
\label{suppl:figscfree}
\end{figure}

To check whether the sharp transition from triangle-poor low-$h$ to triangle-rich high-$h$ state is
a general phenomenon, not a peculiarity of Erd\H{o}s-R\'enyi networks, we have studied the
$\phi(h)$ dependences for the networks with the scale-free node degree distributions of the form
$P(k) \sim k^{-\alpha}$. The networks where prepared via a conventional configurational model
\cite{bender} which is known to give a flat measure over the whole set of networks with a given
degree distribution. Networks with $\alpha = 2,3$ has been studied, which is close to typical
values in experimentally observed scale-free networks, the results are shown in
\fig{suppl:figscfree}. Although the situation is clearly more complicated than in the ER networks
(e.g., the large-$h$ limit of $\phi$ seems to be $N$-dependent while in ER networks it is fully
determined by $p$), the transition clearly does take place. In the case of $\alpha =3$ there seems
to be evidence of an additional intermediate state at $h \sim 0.5$, but this needs further
investigation. Note also that since the networks considered here are relatively small, the heavy
tail of the $P(k)$ distribution is not sufficiently well-represented. Further, more detailed
studies of the localization transition in scale-free networks will be provided elsewhere.

\end{appendix}

\end{document}